# The Dead Cryptographers Society Problem

André Luiz Barbosa
http://www.andrebarbosa.eti.br
Non-commercial projects: SimuPLC – PLC Simulator & LCE – Electric Commands Language

***Abstract***. *This paper defines The Dead Cryptographers Society Problem – DCS (where several great cryptographers created many polynomial-time Deterministic Turing Machines (DTMs) of a specific type, ran them on their proper descriptions concatenated with some arbitrary strings, deleted them and left only the results from those running, after they died: if those DTMs only permute and sometimes invert the bits on input, is it possible to decide the language formed by such resulting strings within polynomial time?), proves some facts about its computational complexity, and discusses some possible uses on Cryptography, such as into distance keys distribution, online reverse auction and secure communication.*

## Contents

## 1. Introduction

In [1], I have promised some practical applications for some new concepts over there introduced to the Theoretical Computer Science community.

In this little paper, I hope I begin to pay that promise.

## 2. The Dead Cryptographers Society Problem

**Definition 2.1. Poly-Time Permuting/Inverting DTM (PTPI DTM).** A DTM (deterministic Turing Machine, or an algorithm or computer program with unbounded memory) *M* is a ***PTPI DTM*** iff its only function is to take a [finite but unbounded] string of **n** input bits and produce from them a string of **n** bits that forms a permutation of the input, and then an arbitrary number of bits, from none to all them, are inverted in order to generate the output: $M(b_1b_2...b_n) = r_{\sigma(1)}r_{\sigma(2)}...r_{\sigma(n)}$, where each **$r_i$** is an individual bit, $1 \leq \sigma(i) \leq n$, $\sigma(i) = \sigma(j) \Leftrightarrow i = j$, and either $r_{\sigma(i)} = b_{\sigma(i)}$ or $r_{\sigma(i)} = \neg b_{\sigma(i)}$. Moreover, a PTPI DTM *M* running on empty (null) string returns a string representing the running time upper bound for that machine *M* ("T($n$) = *1000\*n^2*", for example, where $n$ = |input|).

**Definition 2.2. The Dead Cryptographers Society Problem (DCS).** The *DCS* is an $L_z$-language [2] *L*, or a *promise problem* [9] $\prod$, where the *promise* is the set formed with all words (or binary strings) **w** generated by some PTPI DTM *M* running on binary description or code of another arbitrary PTPI DTM *N* concatenated with some arbitrary binary string **s**: $\prod_{YES}$ ∪ $\prod_{NO} = M(\langle N,s \rangle)$, and $\prod_{YES}$ is the subset formed with all **w** when *N* is the proper binary description or code of that same DTM *M*: $\prod_{YES} = M(\langle M,s \rangle)$. Remember that *M* running on empty string returns a string representing the polynomial running time bound for *M* itself.

**Theorem 2.1. The DCS is in NP.**

*Proof.* See that the **DCS** is in NP [2] (in Promise-NP [9, 11], more precisely), since if **w** ∈ **DCS**, then it is verifiable within polynomial time whether a [guessed] PTPI DTM *M* on an [guessed] input $\langle M,s \rangle$ writes **w** = $\langle M,s \rangle$ on its tape and then halts within polynomial on |w| (*M*("") = string representing $p(|w|)$) steps, since at most $2^{|w|}$ different strings, hence PTPI DTMs, may be coded into **w**, and the running time of that [guessed] *M* is poly-bounded: it suffices to simulate the running of *M* on $\langle M,s \rangle$ during at most that bound of steps and then to check the contents of the tape after the simulation has finished, in order to see whether or not it is equal to $\langle M,s \rangle$. Note that simulating the running of any poly-time DTM on an arbitrary input can be done within polynomial time too (as a polynomial is a time-constructible function [3]). □

**Theorem 2.2. The DCS is not in P.**

*Proof.* Can the **DCS** be in P? No, at all, plainly, since there is no polynomial that bounds the running time of the all PTPI DTM (and the polynomial that bounds the running time of a specific one is not known neither given for us), and then it is impossible to decide or to reduce the **DCS** into another NP problem within deterministic polynomial time.

That is, the Cook-Levin Theorem is false, since in order to all its proofs work a fundamental – but in general neglected – input (information) is absolutely essential and tacitly supposed to be known and given to us (or can be computed in deterministic poly-time) in their scope [2]: the polynomial running time *p(n)* of some NTM (Nondeterministic Turing Machine) that decides an NP problem ($n$ = |input for it|). With such a *p(n)* anyway provided (which can be considered a hidden assumption or axiom), a poly-time DTM that reduces any instance of this problem to an instance of poly-size Boolean formula (or of poly-size Boolean circuit that simulates the problem) can always be constructed by us within deterministic poly-time (where the size of this generated Boolean formula (or circuit) is in $O(log(n)p(n)^2)$).

However, if such a *p(n)* is unknown (or is not given to us) neither can be computed in deterministic polynomial time, then all these proofs shall fail. They stop utterly working, consequently, when such a *p(n)* is unknown (or not given to us) *a priori*, since it cannot be computed in deterministic polynomial time, as widely known.

See [1], [2] and [10] for details of the demonstration of this amazing fact, and the Sections 2.1, 2.2 and 3 below, for some simple examples of possible applications and codifications. □

## 2.1 Online reverse auction mathematically proven poly-unbreakable

We can now construct a protocol for online [virtual] reverse auction [4] mathematically proven unbreakable within polynomial time, by the Theorem 2.2:

It suffices the bidder at the reverse auction to construct a PTPI DTM $M$, to run it on $\langle M,b \rangle$, where **b** is a string representing its bid (into some codification form or standard ruled by the auctioneer), to concatenate the output with $\mathbf{H}(\langle M,M' \rangle)$ (where $M'$ is the inverse machine of that $M$ ($M'(M(x)) = x$) and **H** is some arbitrary cryptographic hash function [6] ruled by the auctioneer too, whom provides it to all the bidders), to store the generated string **w** in a trusted third, sending it to the reverse auctioneer too, and then to hide the PTPI DTMs $M$ and $M'$ from the external world, which assures the confidentiality of that bid. Any crypto hash goes here.

After, in a virtual public session, all the bidders present to the reverse auctioneer their respective DTMs $M$ and $M'$ in order to recuperate their bids (the winner is who has made the lowest one), where obviously there cannot be fraud, because running that $M$ on $M'(M(\langle M,b \rangle))$ during at most the bound of steps (given by that polynomial $M("")$ on $|input|$) and concatenating the output with $\mathbf{H}(\langle M,M' \rangle)$ must result in that same string **w** sent to the auctioneer, and any other input must result in another [different] output $\mathbf{w'} \neq \mathbf{w}$, since $M$ and $M'$ must be PTPI [bijective] machines. Remember yet that $M(\langle M,b \rangle)$ is part of **w**, thus it must be right. Finally, in case of doubts that trusted third can guarantee the authenticity of that **w**.

## 2.2 Distance keys distribution mathematically proven poly-unbreakable

**Definition 2.2.1. n-permuting/inverting DTMs set (n-pi-DTM-set).** An ***n-pi-DTM-set*** is a set of **n** PTPI DTMs $M_1$, $M_2$, …, $M_n$ (not necessarily different machines), where $M_n(M_{n-1}(…(M_2(M_1(\langle M_1,x \rangle)))…)) = \langle M_1,x \rangle$, that is, the result from running $M_1$ on $\langle M_1,x \rangle$, then running $M_2$ on that result, and so on, until $M_n$, is that same original string $\langle M_1,x \rangle$. Notice that a **1-pi-DTM-set** is an identity PTPI DTM $I$ ($I(w) = w$).

With the definition above, we can now also construct a system of distance keys distribution [5] mathematically proven unbreakable into polynomial time, by the Theorem 2.2:

It is enough to utilize a **4-pi-DTM-set** where the user **A** creates a key **k** and then sends **k'** = $M_1(\langle M_1,k \rangle)$ to the user **B**, that returns **k''** = $M_2(k')$ to **A**. **A** then returns **k'''** = $M_3(k'')$ to **B**, that recuperates that original $\langle M_1,k \rangle = M_4(k''')$ and then **k** = right($\langle M_1,k \rangle$; $|\langle M_1,k \rangle|-|\langle M_1 \rangle|$) (**k** is the string formed by the |k| most right positions from $\langle M_1,k \rangle$). See that **B** can eventually check whether or not the messages are authentic, for the DTM $M_1$ from the **4-pi-DTM-set** is sent encoded into each message and it is recovered at final.

Observe that the user **A** does not need to have neither to know $M_2$ and $M_4$, and **B** does not need to have neither to know $M_1$ and $M_3$.

Obs.: notice that, although the users need either trust an initial **4-pi-DTM-set** created by someone else or in-person meeting in order to create that **4-pi-DTM-set**, they need do this only once: all communications thereafter can be made by the established secure channel, even to exchange other **n-pi-DTM-set(s)**.

Verify that eavesdropping those exchange messages with **k'**, **k''** or **k'''** does not allow to recover that original key **k**, since these strings are just permutations/inversions of the bits from $M_1$ concatenated with **k**, and the 4-permuting/inverting machines set used in that exchanging of messages is utterly hidden from everyone else besides **A** and **B**.

Note that this method can be used to send any arbitrary string from **A** to **B**, or from **B** to **A**, even other **4-pi-DTM-set**, in order to avoid utilizing the same one for many times or during a large term, which would be very harmful to the safety of that protocol.

See yet that we can construct a very similar protocol with more general DTMs, not just permuting/inverting ones, even with non-length-preserving DTMs (where |input| ≠ |output|).

### 2.2.1 Example of a 4-pi-DTM-set

Let $f: \{0,1\}^4 \rightarrow \{0,1\}^4$ be a permuting function where:

| (Position of the bit into $x$) = $i$ | $\sigma(i)$ = (Position of that bit into $f(x)$) = $2*i \bmod 5$ |
|---|---|
| 1 | 2 |
| 2 | 4 |
| 3 | 1 |
| 4 | 3 |

Verify that $f(f(f(f(x)))) = x$, hence four equal DTMs $M$ that split the [e.g. ASCII coded] message into pieces or parts of four bits, apply repeatedly that function on each part, and concatenate that outputs, form a **4-pi-DTM-set**, as a consequence of $2^4 \equiv 1 \pmod 5$.

In order to illustrate the functioning of this **4-pi-DTM-set**, we can use it on a little and simple plain message:

| Plain Text | M | | | | A | | | | T | | | | H | | | | | | | | | | | | | | | | | | |
|---|---|---|---|---|---|---|---|---|---|---|---|---|---|---|---|---|---|---|---|---|---|---|---|---|---|---|---|---|---|---|---|
| Parts | **Part$_1$** | | | | **Part$_2$** | | | | **Part$_3$** | | | | **Part$_4$** | | | | **Part$_5$** | | | | **Part$_6$** | | | | **Part$_7$** | | | | **Part$_8$** | | | |
| Bit Pos. | 1 | 2 | 3 | 4 | 1 | 2 | 3 | 4 | 1 | 2 | 3 | 4 | 1 | 2 | 3 | 4 | 1 | 2 | 3 | 4 | 1 | 2 | 3 | 4 | 1 | 2 | 3 | 4 | 1 | 2 | 3 | 4 |
| Bits $x_0$ | 0 | 1 | 0 | 0 | 1 | 1 | 0 | 1 | 0 | 1 | 0 | 0 | 0 | 0 | 0 | 1 | 0 | 1 | 0 | 1 | 0 | 1 | 0 | 0 | 0 | 1 | 0 | 0 | 1 | 0 | 0 | 0 |
| $x_1 = f(x_0)$ | 0 | 0 | 0 | 1 | 0 | 1 | 1 | 1 | 0 | 0 | 0 | 1 | 0 | 0 | 1 | 0 | 0 | 0 | 1 | 1 | 0 | 0 | 0 | 1 | 0 | 0 | 0 | 1 | 0 | 1 | 0 | 0 |
| 'Text' $x_1$ | | ☻ | | | | | | | | ☻ | | | | | | | | | 1 | | | | | | | ☻ | | | | | | |
| $x_2 = f(x_1)$ | 0 | 0 | 1 | 0 | 1 | 0 | 1 | 1 | 0 | 0 | 1 | 0 | 1 | 0 | 0 | 0 | 1 | 0 | 1 | 0 | 0 | 0 | 1 | 0 | 0 | 0 | 1 | 0 | 0 | 0 | 0 | 1 |
| 'Text' $x_2$ | | + | | | | | | | | ( | | | | | | | | Ó | | | | | | | | ! | | | | | | |
| $x_3 = f(x_2)$ | 1 | 0 | 0 | 0 | 1 | 1 | 1 | 0 | 1 | 0 | 0 | 0 | 0 | 1 | 0 | 0 | 1 | 1 | 0 | 0 | 1 | 0 | 0 | 0 | 1 | 0 | 0 | 0 | 0 | 0 | 1 | 0 |
| 'Text' $x_3$ | | Ä | | | | | | | | Ä | | | | | | | | Ã | | | | | | | | É | | | | | | |
| $x_4 = f(x_3)$ | 0 | 1 | 0 | 0 | 1 | 1 | 0 | 1 | 0 | 1 | 0 | 0 | 0 | 0 | 0 | 1 | 0 | 1 | 0 | 1 | 0 | 1 | 0 | 0 | 0 | 1 | 0 | 0 | 1 | 0 | 0 | 0 |
| Text $x_4$ | | **M** | | | | | | | | **A** | | | | | | | | **T** | | | | | | | | **H** | | | | | | |

Note that replacing above those $\{0,1\}^4$ by $\{0,1\}^{p-1}$ (where $p$ is any odd prime number), that formula $2*i \bmod 5$ by $k*i \bmod p$, where $k$ is not multiple of $p$, and "*groups of four bits*" by "*groups of p-1 bits*", we can construct in a similar manner an $ord_p(k)$-**pi-DTM-set**, because $k^{ord_p(k)} \equiv 1 \pmod p$, that is, the bits return to their original positions after $ord_p(k)$ $M$ runnings, where $ord_p(k)$ is *the multiplicative order of k modulo p*. [7]

Notice yet that any **n-pi-DTM-set** whose construction is publicly known shall be very unsafe for cryptographic purposes, obviously. So, they must be constructed with creativity and originality, if they are intended to be used into some serious cryptographic protocol.

## 2.3 Secure communication mathematically proven poly-unbreakable

With the protocol introduced in Section 2.2, we can make a secure communication mathematically proven poly-unbreakable, by the Theorem 2.2: it suffices to utilize a **2-pi-DTM-set** where a user **A** can send an arbitrary message **m** by sending **m'** = $M_1(\langle M_1, m \rangle)$ to the user **B**, that recuperates that original $\langle M_1, m \rangle = M_2(m')$ and then **m** = right($\langle M_1, m \rangle$; $|\langle M_1, m \rangle| - |\langle M_1 \rangle|$) (**m** is the string formed by the |m| most right positions from $\langle M_1, m \rangle$).

As in Section 2.2, **B** can eventually check the authenticity of the messages and this method can even be used to a user send other **2-pi-DTM-set** to another one, in order to avoid using the same DTMs repeatedly, by safety reasons. If they want it, the user **A** can yet send only *m* to **B**, without concatenating with ⟨$M_1$⟩ neither with something else at all, or concatenating it with any other arbitrary string(s).

Observe also that the Kerckhoffs' Principle [12] is not valid for the new cryptographic processes introduced herein, since the proper *n-pi-DTM-sets* act as were the *keys* w.r.t these processes.

## 3. Example of a 2-pi-DTM-set in JavaScript

The JavaScript code *f* below is an example of a **2-pi-DTM-set**, since it is a PTPI DTM and $f(f(x)) = x$:

```
<textarea id="ti1" value="TEST" rows="15" cols="190">Enter herein the encrypting/decrypting code

(Example of code on the paper "The Dead Cryptographers Society Problem",
at https://arxiv.org/ftp/arxiv/papers/1501/1501.03872.pdf)
{
Enter herein the plain/cyphered text in order to be cyphered/decyphered
}
</textarea>
<textarea id="ti2" value="TEST" rows="5" cols="190">That plain/cyphered text showed in binary</textarea>
<textarea id="ti3" value="TEST" rows="5" cols="190">The cyphered/decyphered text showed in binary</textarea>
<textarea id="ti4" value="TEST" rows="15" cols="190">That cyphered/decyphered text showed in character</textarea>
<textarea id="ti5" value="TEST" rows="1" cols="190"></textarea>
<button onclick="Convert();">Encrypt/Decrypt</button>

<script>
function Convert() {
  var Plain_Text    = document.getElementById("ti1"); // The plain/cyphered text in ASCII
  var Binary_Text   = document.getElementById("ti2"); // The plain/cyphered text into binary
  var Binary_Cypher = document.getElementById("ti3"); // The cyphered/decyphered text into binary
  var Cypher_Text   = document.getElementById("ti4"); // The cyphered/decyphered text in ASCII
  var Msg_Text      = document.getElementById("ti5"); // The msg text for copy to the clipboard

  if (Plain_Text.value == ""){
   Cypher_Text.value = "T(n) = 1000*n^2";
  }
  else{
   var p = 1288; // Size of the blocks (should be an odd prime - 1)
   var q = 3456; // Controls 1 and 2 of the bit-inverted positions ...
   var r = 7890; // ... (can be any two arbitrary integers)

   k = Plain_Text.value.length;
   Binary_Text.value = "";

   for (i=0; i < k; i++) { // That plain/cyphered text showed in binary
     Binary_Text.value += Right("00000000" + Plain_Text.value[i].charCodeAt(0).toString(2), 8);
   }

   Binary_Cypher.value = Binary_Text.value;
   k = Binary_Text.value.length;
   s = k % p;
   k -= s;

   for (i=0; i < k; i++) { // The cyphered text transformed and showed in binary
     j = (i % p) + 1;
     Perm_i = (p*j) % (p+1) - 1 + parseInt(i/p) * p;
     if ((q % ((q*r)/(Perm_i+i))) % 2)
       Binary_Cypher.value = setCharAt(Binary_Cypher.value, i, Binary_Text.value.charAt(Perm_i));
     else
       Binary_Cypher.value = setCharAt(Binary_Cypher.value, i, Binary_Text.value.charAt(Perm_i) == "0" ? "1" : "0");
   }

   k += s;
   for (i=k-s; i < k; i++) { // The cyphered text transformed and showed in binary
     Binary_Cypher.value = setCharAt(Binary_Cypher.value, i, Binary_Text.value.charAt(i) == "0" ? "1" : "0");
   }

   Cypher_Text.value = "";
   k--;
   for (i=0; i < k; i += 8) { // The cyphered text transformed and showed in character
```

```
      Cypher_Text.value += String.fromCharCode(parseInt(Binary_Cypher.value.substring(i, i+8), 2)); }
    document.getElementById("ti4").focus();
    document.getElementById("ti4").select();
    Msg_Text.value = "<Ctrl + C> Copies the Cyphered/Decyphered Text to the Clipboard";
  }
}
function Right(str, n){
  if (n <= 0)
    return "";
  else if (n > String(str).length)
    return str;
  else {
    var iLen = String(str).length;
    return String(str).substring(iLen, iLen - n);
  }
}
function setCharAt(str, index, chr) {
  if(index > str.length-1) return str;
  return str.substring(0, index) + chr + str.substring(index+1);
}
function gcd(a, b) {
  return !b ? a : gcd(b, a % b);
}
</script>
```

The code above can be run and tested in this webpage:

http://www.andrebarbosa.eti.br/Cryptor-DeCryptor_DCS.html.

Into this webpage, enter (or copy/paste) a plain text in order to be ciphered in the first field, press the <Encrypt/Decrypt> button, and then the ciphered text shall appear (selected) in the result field. Copy this ciphered text, paste it in the first field and press again the <Encrypt/Decrypt> button. Then, the original plain text should come out deciphered in the result field.

Notice that, even though trillions of different **2-pi-DTM-set**'s can be created basing on that code simply by mean of changing the integer variables **p**, **q** and **r** into it, it is very unsecure using it in order to seriously encrypt messages or files, since that code is publicly known. However, verify whether that security of this encryption can be improved if the message or file contents (plain text) is padded with **(LCM(p, 8\*length of it** [in characters]**) – 8\*length of it** [in characters]**)** repeated characters: "**\***", for example. (Why or why not?)

Challenge for cryptanalysis: The integers **p**, **q** and **r** into the JavaScript code above were changed, thereby generating a new specific **2-pi-DTM-set**. Then, a small message was encrypted by it, resulting the following ciphered text (some characters are non-printable, but they can be read as ASCII bits below):

    YéÕûKY11Yû   ûY9‰ÑyMû™ÑyIYéiû19iY±  û1‹I      ÙY

(Into ASCII bits:    01011001 11101001 11010101 11111011 01001011 01011001 00110001 00110001 01011001 11111011 00001001 11111011 01011001 00111001 10001001 11010001 01111001 01001101 11111011 10011001 11010001 01111001 01001001 01011001 11101001 01101001 11111011 00110001 00111001 01101001 01011001 10110001 10011101 11111011 00110001 10001011 01001001 00001001 11011001 01011001)

The challenge is: What is the plain text of that message? Is there an efficient algorithm in order to compute those integers **p**, **q** and **r** (given that JavaScript code and arbitrary number of ciphered texts of arbitrary lengths)?

## 4. Conclusion

The conclusion is that we can create very secure communication protocol, like other surprising algorithms and programs, using disparaged ideas; hence maybe we should not summarily disregard new ideas into Theoretical Computer Science as weird and irrelevant ones, because they can perhaps be useful in unexpected applications upon inventive and creative minds, into a very dynamical technological world.

## 5. Freedom & Mathematics

"**– The essence of Mathematics is Freedom.**" (Georg Cantor) [8]

## 6. References

[1]  A. L. Barbosa, *The Cook-Levin Theorem is False*, unpublished, available: http://www.andrebarbosa.eti.br/The_Cook-Levin_Theorem_is_False.pdf

[2]  A. L. Barbosa, *P != NP Proof*, unpublished, available: http://arxiv.org/ftp/arxiv/papers/0907/0907.3965.pdf

[3]  From Wikipedia, the free encyclopedia, "*Constructible Function*", unpublished, available: http://en.wikipedia.org/wiki/Constructible_function

[4]  From Wikipedia, the free encyclopedia, "*Online Auction*", unpublished, available: http://en.wikipedia.org/wiki/Online_auction

[5]  From Wikipedia, the free encyclopedia, "*Key Distribution Center*", unpublished, available: http://en.wikipedia.org/wiki/Key_distribution_center

[6]  From Wikipedia, the free encyclopedia, "*Cryptographic Hash Function*", unpublished, available: http://en.wikipedia.org/wiki/Cryptographic_hash_function

[8]  From The Engines of Our Ingenuity, site, "*Episode nº 1484: GEORG CANTOR*", posted by John H. Lienhard, unpublished, available: http://www.uh.edu/engines/epi1484.htm

[9]  O. Goldreich, *On Promise Problems (in memory of Shimon Even (1935-2004))*, unpublished, available: http://www.wisdom.weizmann.ac.il/~oded/PS/prpr.ps

[10] A. L. Barbosa, *NP $\not\subset$ P/poly Proof*, unpublished, available: http://www.andrebarbosa.eti.br/NP_is_not_in_P-Poly_Proof_Eng.pdf

[11] L. A. Hemaspaandra, K. Murray, and X. Tang, *Barbosa, Uniform Polynomial Time Bounds, and Promises*, Technical Report, unpublished, available: http://arxiv.org/abs/1106.1150

[12] From Wikipedia, the free encyclopedia, "*Kerckhoffs' Principle*", unpublished, available: https://en.wikipedia.org/wiki/Kerckhoffs%27s_principle